\documentclass[12pt]{revtex4}

\usepackage{graphicx}        
\usepackage{grffile}         
\usepackage{graphics}
\usepackage{epsfig}
\usepackage{verbatim}
\usepackage{dcolumn}        
\usepackage{bm}            
\usepackage{amsmath}
\usepackage{mathtools}
\usepackage{physics}
\usepackage{hyperref}        
\hypersetup{colorlinks,breaklinks,
            linkcolor=black,urlcolor=black,
            anchorcolor=black,citecolor=black}
\usepackage{color}
\usepackage{float}
\newcommand{\EQ}[1]{Eq.~(\ref{eq:#1})}

\def\bal#1\eal{\begin{align}#1\end{align}}
\newcommand{\be}{\begin{equation}}
\newcommand{\ee}{\end{equation}}
\newcommand{\ba}[1]{\begin{array}{*{#1}{c}}}
\newcommand{\ea}{\end{array}}

\newcommand{\old}{\color{black}} 

\graphicspath{{./Figures/}} 

\begin{document}
\title{Solution of the Metropolis dynamics on a complete graph with application to the Markov chain Mpemba effect}

\author{Israel Klich and Marija Vucelja}
\affiliation{$^{1}$ Department of Physics, University of Virginia, Charlottesville, VA 22904, USA}
\email{klich@virginia.edu}
\email{mvucelja@virginia.edu}

\begin{abstract}
We find analytically the complete set of eigenvalues and eigenvectors associated with Metropolis dynamics on a complete graph. As an application, we use this information to study a counter-intuitive relaxation phenomenon, called the Mpemba effect. This effect describes situations when upon performing a thermal quench, a system prepared in equilibrium at high temperatures relaxes faster to the bath temperature than a system prepared at a temperature closer to that of the bath. We show that  Metropolis dynamics on a complete graph does not support weak nor strong Mpemba effect, however, when the graph is not complete, the effect is possible.
\end{abstract}


\maketitle

\section{Introduction}
Monte Carlo algorithms are widely used in diverse fields of science,~\cite{97Sokal,Kra06}. In particular, they are often the tool of choice when estimating multidimensional integrals present in statistical physics and quantum field theory. Typically in physical systems, describing the real world, the phase space is exponentially large. Monte Carlo algorithms are especially useful for sampling equilibrium properties, and estimating equilibrium observables, where direct sampling is unfeasible due to an exponentially large phase space. Their convergence properties are widely studied, see, e.g.~\cite{2009Levinbook}.

In a typical implementation for a system with $n$ states, one specifies the probability distribution $\ket{p} = (p_1, p_2,...,p_n)$ where $p_i(t)$ is the probability for the system to be in state $i$ at time $t$. The evolution of the system is assumed to depend only on the present, i.e. it is Markovian, and it obeys a Master equation 
\bal 
\label{eq:MasterEq}
\partial _t \ket{p}= R \ket{p}, 
\eal 
where the off-diagonal matrix element $R_{ij}$ is the rate (probability per unit time) to jump from state $j$ to $i$. The diagonal elements of the rate matrix are "escape rates" 
\bal
R_{ii}  = - \sum _{j; j \neq i} R_{ji},
\eal
set so that the sum over any column of $R$ gives zero. This constraint ensures the conservation of probability. For a general treatment, see  e.g.~\cite{97Sokal,2009Levinbook}. 
We will assume that under the Markov evolution, the system converges to a steady state, $\ket{\pi}$. The global balance condition guarantees that total flux from any of state of the systems is equal to the total influx to that state, or
\bal
\label{eq:GBcond}
\sum_j R_{ij} \pi_j = \sum _j R_{ji} \pi_i, \quad \forall i. 
\eal 
Typically, such a non-local condition is hard to implement in practice, except for select cases, see, e.g. ~\cite{turitsyn2011irreversible,LiftingVucelja}. Detailed balance is a special case of \EQ{GBcond}
\bal
R_{ij} \pi_j = R_{ji}\pi_i,  \quad \forall i, j. 
\eal
 Despite often slow relaxations to a steady state~\cite{turitsyn2011irreversible,LiftingVucelja}, Markov jump processes obeying detailed balance, are preferred, as detailed balance is an easily implementable, local condition. In particular, one of the most popular Monte Carlo jump process dynamics, that obeys detail balance is the Metropolis dynamics ~\cite{MRRTT53}. Metropolis dynamics is described by the following transition acceptance rate from state $j$ to state $i$:
\bal 
R_{ij} = {\rm min}\left(1, \frac{\pi _i}{\pi _j}\right)\label{Metropolis_rule}.
\eal
Note that it is always possible to view the steady state as a thermal equilibrium, by writing $\pi_i \propto e^{-E_i/k_BT}$, i.e.~the Bolzmann distribution at some temperature $T$ and $\{E_i\}$ would correspond to energies. A particularly simple choice is $E_i=-\log{\pi_i}$, which sets $\beta=1$ and $Z=1$.

In this paper, we solve for the eigensystem of Metropolis dynamics for a complete graph, as described in the next section. We then use the slowest relaxing eigenvector to explore the possibility of a Mpemba effect (defined below) in such a system.  

\section{Metropolis dynamics on a complete graph}

When only transitions between specific pairs of states $(i,j)$ appear in the rate matrix $R$, we say that the  Metropolis  dynamics takes place on a graph corresponding to the adjacency matrix $A$, where $A_{ij}=1$ if $(i,j)$ is allowed, $A_{ij}=0$ if $(i,j)$ is not allowed, and set $R_{ij}$=0. The rule \eqref{Metropolis_rule}, without further restriction on which pairs of states $j$ and $i$ admit transitions, describes dynamics on a complete graph. Otherwise, if only transitions between certain pairs of states $(i,j)$ are allowed directly in $R$, the graph is not complete. For example for a spin system, one often studies the Glauber dynamics, where only states differing by a single spin state are connected~\cite{Glauber}.

Due to the appearance of the $min$ in the form of the transition matrix \eqref{Metropolis_rule}, the matrix is non-analytic as a function of the energy levels. Despite this non-analyticity, we find, somewhat surprisingly, that the entire spectral decomposition of $R$ is given in a simple form. The results are expressed in the following theorem below.\old

We will use the standard basis vectors  $|1\rangle ,..,|n\rangle $, where $\ket{j} = (0,0,...,0,1,0,...,0)$ is an $n-$dimensional vector, with $1$ is in the $j-$th place.

\emph{Theorem:} Let an ordering of the steady state probabilities be given by a permutation $r:\{1,..,n\}\rightarrow \{1,..,n\}$, s.t. $\pi_{r_1}\geq \pi_{r_2}...\geq \pi_{r_n}$, or, equivalently, the energies corresponding to the steady state distribution $\pi$ are ordered as 
\bal
E_{r_1} \leq E_{r_2} \leq ... \leq E_{r_n}.
\eal
Let the permutation matrix realizing the permutation $r$ be $S_r$, s.t. $S_r\ket{i}=\ket{r_i}$
The first eigenvalue and its associated eigenvector of the Metropolis $R$ are 
\bal
&\lambda_1 = 0,~~ \quad \bm |v_1\rangle = (\pi_1,\pi_2 , ..., \pi_n).
\eal
The rest of the eigenvalues, ordered by absolute value, are given by ${S_r}|v_k\rangle$, where
\bal
& |v_k\rangle = ((k - 2) \text{ zeros}, - Z_k , \pi_{r_k} , ..., \pi_{r_n}), ~~1<k\leq n,
\\
& \lambda_k = - \left(\pi_{r_{k-1}}^{-1}Z_{k-1} + k -2\right),
\eal
are the eigenvectors of the system in the basis where the $\pi_i$ are ordered by magnitude. 
Here
\bal
&Z_k \equiv \sum ^n _{l = k} \pi_{r_l}.
\eal

{\it Proof:}
Below for convenience we chose to work with the Boltzmann form of the $\pi_i$ instead of directly with $\pi_i$. 
First, we note that for any ordering of the energies, we can use the transformation $R \rightarrow (S_r)^{-1} R S_r$ to first work in a basis where the energies are ordered according to their magnitude. It will therefore be enough to prove the relation under the assumption that the energy levels are ordered as $E_1\leq E_2 \leq ... \leq E_n$ at the outset. In the end we just return to the original ordering by applying ${S_r}$ to the resulting eigenvectors.
We will now do induction on the number of energy levels. 
 Our induction assumption is that for $1<k\leq n$,
\begin{equation}
 |v_k\rangle =- Z_k|k-1\rangle +\Sigma _{l=k}^ne^{-\beta E_l}|l\rangle
\end{equation}
are the eigenstates of $R$ with corresponding eigenvalues $\lambda _k=-\left(e^{\beta  E_{k-1}}Z_{k-1}+k-2\right)$. 
In addition we have the vector 
\begin{equation}
 |v_1\rangle =\Sigma _{l=1}^n e^{-\beta 
   E_l}|l\rangle,
\end{equation}
as a zero eigenvector by construction of the  Metropolis  matrices.
We first check our initial assumptions on a $2\times 2$  Metropolis system, with energies $E_1<E_2$. The $R$ matrix is:
\begin{equation}
    \left(
\begin{array}{cc}
 -e^{\beta  E_1}e^{-\beta  E_2} & 1 \\
 e^{\beta  E_1}e^{-\beta  E_2} & -1 \\
\end{array}
\right)
\end{equation}
This system has the eigenvectors and eigenvalues:
\bal
&
|v_1\rangle =\left(e^{-\beta  E_1},e^{-\beta E_2}\right)\text{  };\text{   }\lambda _1=0,  
\\ 
& |v_2\rangle =\left(-Z_2,e^{-\beta 
   E_2}\right)\text{  };\text{   }\lambda
   _2=-Z_1e^{\beta  E_1}.
\eal
These eigenvalues fit the desired form.

We will assume that the form is correct for levels $E_1\leq E_2\leq E_3 \leq ...\leq E_n$, and then consider adding another energy level $E_0\leq E_1$ to the complete graph.
With the energy ordering chosen, the $R$ matrix has the form:
\begin{eqnarray*} & R_{[1;n]}={\tiny\left(
\begin{array}{cccccc}
 -\Sigma _{k=2}^n e^{-\beta 
    (E_k-E_1 )} & 1 & 1 & 1 & 1 & 1
   \\
 e^{-\beta   (E_2-E_1 )} & -1-\Sigma
   _{k=3}^n e^{-\beta   (E_k-E_2 )} &
   1 & 1 & 1 & 1 \\
 e^{-\beta   (E_3-E_1 ) } & e^{-\beta 
    (E_3-E_2 ) } & \text{..} &
   \text{..} & 1 & 1 \\
 e^{-\beta   (E_4-E_1 ) } & e^{-\beta 
    (E_4-E_2 ) } & \text{..} &
   \text{...} & 1 & 1 \\
 \text{...} & \text{...} & \text{...} &
   \text{...} & -(n-2)-e^{-\beta 
    (E_n-E_{n-1} ) } & 1 \\
 \text{...} & \text{...} & \text{...} &
   e^{-\beta   (E_n-E_{n-2} ) } &
   e^{-\beta   (E_n-E_{n-1} ) } &
   -(n-1) \\
\end{array}
 \right)}
 \\ & ={\tiny\left(
\begin{array}{cccccc}
 -e^{\beta  E_1}Z_2 & 1 & 1 & 1 & 1 & 1 \\
 e^{-\beta    (  E_2-E_1 ) } &
   -1-e^{\beta  E_2}Z_3 & 1 & 1 & 1 & 1 \\
 e^{-\beta    (  E_3-E_1 ) } & e^{-\beta 
     (  E_3-E_2 ) } & \text{..} &
   \text{..} & 1 & 1 \\
 e^{-\beta    (  E_4-E_1 ) } & e^{-\beta 
     (  E_4-E_2 ) } & \text{..} &
   \text{...} & 1 & 1 \\
 \text{...} & \text{...} & \text{...} &
   \text{...} & -(n-2)-e^{-\beta 
     (  E_n-E_{n-1} ) } & 1 \\
 \text{...} & \text{...} & \text{...} &
   e^{-\beta    (  E_n-E_{n-2} ) } &
   e^{-\beta    (  E_n-E_{n-1} ) } &
   -(n-1) \\
\end{array}
\right)}.
 \end{eqnarray*}

Now we add a new energy $E_0\leq E_1$. We  write the new $(n+1)\times (n+1)$ matrix $R_{[0;n]}$ as:
 \begin{equation}
     R_{[0;n]}={ \left(
\begin{array}{cccc}
 -e^{\beta  E_0}Z_1  & 1 & \text{..} & 1 \\
 e^{\beta  E_0}e^{-\beta  E_1} &
   \text{..} & \pmb{\text{..}} &
   \pmb{\text{..}} \\
 \text{..} & \pmb{\text{..}} & \pmb{R_{[1;n]}}
   & \pmb{\text{..}} \\
 e^{\beta  E_0}e^{-\beta  E_n} &
   \pmb{\text{..}} & \pmb{\text{..}} &
   \pmb{\text{..}} \\
\end{array}
\right)-\left(
\begin{array}{cccc}
 0 & 0 & 0 & 0 \\
 0 & 1 & 0 & 0 \\
 0 & 0 & \text{..} & 0 \\
 0 & 0 & 0 & 1 \\
\end{array}
\right)},
 \end{equation}
 where the lower right $n\times n$ block in the first term is simply $R_{[1;n]}$. Below we will be using the natural basis elements labeled as: $|0\rangle ,|1\rangle ,...,|n\rangle $.  We will define: 
\begin{align}
R_{[1;n]}\equiv\Sigma_{i,j=1}^nR_{\text{ij}}|i\rangle \langle j|
\end{align}
and 
\begin{align}
P_{[1;n]}\equiv\Sigma
   _{i=1}^n |i\rangle \langle i|~~.
\end{align}
With these definitions
 \begin{equation}
R_{[0;n]}=R_{[1;n]}-P_{[1;n]}-e^{\beta  E_0}Z_1|0\rangle
   \langle 0|+\Sigma _{l=1}^n|0\rangle \langle l|+\Sigma
   _{l=1}^ne^{\beta  E_0}e^{-\beta  E_l}|l\rangle \langle
   0|.
 \end{equation}
To compare with our induction assumptions, we will label the eigenvectors of the new $(n+1)$ state system $|\tilde{v}_k \rangle$, for $k=1,...,(n+1)$, and we will use the correspondences $\tilde{E}_1=E_0\text{  },\text{  
   }\tilde{E}_2=E_1 ,\text{...} ,\text{ 
   }\tilde{E}_{n+1}=E_n$ for the energies and  $|\tilde{1}\rangle=|0\rangle,|\tilde{2}\rangle=|1\rangle,..,|\tilde{n+1}\rangle=|n\rangle$ for the vectors.
In addition, we now have:
   $\tilde{Z}_k\equiv \Sigma _{l=k}^{n+1} e^{-\beta
    \tilde{E}_l}$.
Note that the above definition implies the relation: $\tilde{Z}_k=Z_{k-1}$.

We now proceed with the induction, separating into the following cases:

(i) $k=1$.
By the construction of  Metropolis, we already know that 
\begin{align}
|\tilde{v}_1\rangle =\Sigma _{l=0}^ne^{-\beta 
   E_l}|l\rangle=\Sigma _{l=1}^{n+1} e^{-\beta 
   \tilde{E}_l}|\tilde{l}\rangle
\end{align}
is the eigenvector with eigenvalue zero.

(ii) $2<k\leq (n+1)$. 
Let us check that indeed $|\tilde{v}_k\rangle $ are the desired eigenstates. First, notice that:
\begin{eqnarray}
& |\tilde{v}_k\rangle=- \tilde{Z}_k|\widetilde{k\!-\!1}\rangle +\Sigma _{l=k}^{n+1} e^{-\beta \tilde{E}_l}|\tilde{l}\rangle=
-Z_{k-1}|(k-2)\rangle +\Sigma _{l=k}^{n+1} e^{-\beta E_{l-1}}|l-1\rangle=|v_{k-1}\rangle.
\end{eqnarray}
We now explicitly apply $R_{[0;n]}$:
\begin{align}
\nonumber
R_{[0;n]}|\tilde{v}_k\rangle
   &=\left(R_{[1;n]}-P_{[1;n]}-e^{\beta 
   E_0}Z_1|0\rangle \langle 0|+\Sigma
   _{l=1}^n|0\rangle \langle l|+\Sigma
   _{l=1}^ne^{\beta  E_0}e^{-\beta 
   E_l}|l\rangle \langle
   0|\right)|v_{k-1}\rangle 
   \\
   & 
   \nonumber
   =\left(\lambda
   _{k-1}-1\right)|v_{k-1}\rangle
   +\left(-e^{\beta  E_0}Z_1|0\rangle \langle
   0|+\Sigma _{l=1}^n|0\rangle \langle
   l|+\Sigma _{l=1}^ne^{\beta  E_0}e^{-\beta 
   E_l}|l\rangle \langle
   0|\right)|v_{k-1}\rangle \\ & =\left(\lambda
   _{k-1}-1\right)|v_{k-1}\rangle, 
\end{align}
where we have used the induction hypothesis:
\begin{align}
R_{[1;n]}|v_{k-1}\rangle =\lambda_{k-1}|v_{k-1}\rangle,    
\end{align}
\begin{align}
\left\langle 0\left|v_{k-1}\right.\right\rangle=0
\end{align}
and that for $k>2$ 
\begin{eqnarray}
\Sigma _{l=1}^n\left\langle
   l\left|v_{k-1}\right.\right\rangle =-
   Z_{k-1}+\Sigma _{l=k-1}^n e^{-\beta E_l}=0. 
   \end{eqnarray}
Finally,  using $\tilde{Z}_k=Z_{k-1},\tilde{E}_k=E_{k-1}$ we have:
\begin{eqnarray}
R_{[0;n]}|\tilde{v}_k\rangle
   =\left(-\left(e^{\beta 
   E_{k-2}}Z_{k-2}+k-3\right)-1\right)|v_{k-1}\rangle =-\left(e^{\beta 
   \tilde{E}_{k-1}}\tilde{Z}_{k-1}+k-2\right)|\tilde{v}_k\rangle,
   \end{eqnarray}
showing that $|\tilde{v}_k\rangle$ is of the desired form.
\\
It is left to check the form for $k=2$. We write:
\begin{eqnarray}
|\tilde{v}_2\rangle =- \tilde{Z}_2|0\rangle
   +\Sigma _{l=1}^ne^{-\beta  E_l}|l\rangle =-
   \tilde{Z}_2|0\rangle +|v_1\rangle.
      \end{eqnarray}
Applying $R$ explicitly, we have:
\begin{align}
\nonumber
R_{[0;n]}|\tilde{v}_2\rangle
   &=\left(R_{[1;n]}-P_{[1;n]}-e^{\beta 
   E_0}Z_1|0\rangle \langle 0|+\Sigma
   _{l=1}^n|0\rangle \langle l|+\Sigma
   _{l=1}^ne^{\beta  E_0}e^{-\beta 
   E_l}|l\rangle \langle 0|\right)(|v_1\rangle-
   \tilde{Z}_2|0\rangle) 
   \\
   \nonumber
   &= -
   \tilde{Z}_2\left(-e^{\beta  E_0}Z_1|0\rangle
   +\Sigma _{l=1}^ne^{\beta  E_0}e^{-\beta 
   E_l}|l\rangle \right)+\left(-|v_1\rangle
   +|0\rangle Z_1\right)
   \\ 
   \nonumber
   &= \left(-
   \tilde{Z}_2e^{\beta 
   E_0}-1\right)\left(-Z_1|0\rangle
   +|v_1\rangle \right)
   \\
   &=-\left(
   \tilde{Z}_2e^{\beta 
   \tilde{E}_1}+1\right)|\tilde{v}_2\rangle
\end{align}
Finally we note that: 
\begin{align}\tilde{Z}_2e^{\beta  \tilde{E}_1}+1=e^{\beta 
   \tilde{E}_1}\Sigma _{l=1}^{n+1}e^{\beta 
   \tilde{E}_l}=e^{\beta 
   \tilde{E}_1}\tilde{Z}_1
\end{align}
giving:
   \begin{align}
   R_{[0;n]}|\tilde{v}_2\rangle =-e^{\beta 
   \tilde{E}_1}\tilde{Z}_1|\tilde{v}_2\rangle,
   \end{align}
which is consistent with: $\tilde{\lambda }_k=\left(e^{\beta 
   \tilde{E}_{k-1}}\tilde{Z}_{k-1}+k-2\right)$ at $k=2$.
This concludes our proof. 

In the next section, we define an interesting counter-intuitive relaxation behavior called the Mpemba effect. Afterward, we apply our results for the Metropolis dynamics, to study the existence of the Mpemba effect for Metropolis dynamics on a complete graph. 

\section{Relaxation with Metropolis dynamics and the Mpemba effect}


The Mpemba effect is a surprising relaxation phenomenon known already since Aristotle times~\cite{Aristotle}. The effect is encountered when a hot liquid relaxes faster to equilibrium than a cold one when they are coupled with a cold bath~\cite{Mpemba}. The effect has been observed several systems, including clathrate hydrates \cite{paper:hydrates}, polymers \cite{18Mpembapoly}, magnetic alloys \cite{Magnetic_Mpemba}, carbon nanotube resonators \cite{Theory_Carbon_nano_greaney2011mpemba}, granular gases \cite{Granular_Mpemba_PhysRevLett} as well, as in a dilute atomic gas \cite{keller2018quenches}.

A phenomenological description of such behavior for Markovian dynamics was recently given in~\cite{2016LuRaz,Mpemba17VRHK}. This approach allows for the investigation of Mpemba like behavior in many different systems as Markovian dynamics is an excellent description of many processes in physics and chemistry (see, e.g. ~\cite{van_Kampen}).

To describe the effect, we assume the relaxation is governed by a master equation with a transition rate matrix $R$ as described above. In the Mpemba effect case, we take the initial condition for \EQ{MasterEq} the thermal equilibrium at some temperature $T\neq T_{b}$, where $T_b$ is the bath temperature:
\bal
p_i(T;t=0)={\pi _i(T)}\equiv \frac{e^{-\beta E_i}}{Z(T)}.
\eal
During the relaxation process, the distribution $\ket{p}$ --- i.e. the solution of \EQ{MasterEq} -- is
\bal
\label{eq:ProbExpansion}
\ket{p(T;t)} = e^{Rt}\ket{\pi(T)}= \ket{\pi(T_b)} + \sum _{i > 1} a_i(T) e^{\lambda_{i}t} \ket{v_i},
\eal 
where the rate matrix $R$ has (right) eigenvectors $\ket{v_i}$ and eigenvalues $\lambda_i$. The largest eigenvalue of $R$, $\lambda _1 = 0$, is associated with the stationary (equilibrium) distribution $\ket{\pi(T_b)}$, whereas all the other eigenvalues have negative real parts, and they correspond to the relaxation rates of the system. The equilibration timescale is typically characterized by $-(\Re\lambda_{2})^{-1}$~\footnote{For detailed balance matrices $R$, the eigenvalues are all real.}.

\section{Existence of Mpemba effects with Metropolis dynamics}
We are now in position to utilize our established result for the second eigenvector of  Metropolis  dynamics on a complete graph. The presence of a Mpemba effect in a relaxation process like Eq. \eqref{eq:MasterEq} can be characterized as follows \cite{2016LuRaz}.  Assume that  $|\Re\lambda_2| < |\Re\lambda_3|$ ($\lambda_k$ are actually real for the  Metropolis  system), then at long times the probability distribution (\ref{eq:ProbExpansion}) is approximated as 
\bal 
\label{eq:p_long_time_limit}
\ket{p(T;t)}\approx \ket{\pi(T_b)} + a_2(T)e^{\lambda_2t}\ket{v_2}.
\eal 
We say that an Mpemba effect exists in this system if there are three temperatures $T_h> T_c> T_b$, such that 
\bal
|a_2(T_h)| < |a_2(T_c)|.
\eal
In other words at long times, the system prepared at $T=T_h$ is closer to the bath equilibrium distribution than a system prepared at $T_c$, although $T_c$ is closer to the bath temperature $T_b$. In \cite{Mpemba17VRHK} a stronger variant of the effect was proposed, where at some temperature $T\neq T_b$, we have $a_2(T)=0$, signaling a jump in relaxation times (${1\over \lambda_2}\rightarrow {1\over \lambda_3}$). An analogues "inverse" Mpemba effect is also possible with $T_c< T_h< T_b$. We summarize all the above cases as in the following definition: 

{\it We say that a system has a strong Mpemba effect if there exists a temperature $T \neq T_{bath}$ for which $a_2(T)=0$, and a weak Mpemba effect if $a(T)$ is a  non-monotonic function  on either $(0,T_b)$ or $ (T_b,\infty)$ (or on both).}

In the next section we show that in our system of Metropolis on a complete graph, both the weak and strong Mpemba effects are absent. Finally, we show a $3\times 3$ example where these effects are possible for Metropolis on a graph which is incomplete.

\subsection{Absence of Mpemba effect with  Metropolis  on a complete graph}
We will use the following expression for $a_2$, derived in \cite{2016LuRaz}, 
\begin{eqnarray}
a_2(\beta)=c(\beta_b) \langle v_2|\mathcal{F} \left(\beta _b\right)|\pi (\beta )\rangle,
 \label{a2_formula}  \end{eqnarray}
where $\mathcal{F} \left(\beta _b\right)$ is a diagonal matrix with elements $\left(\mathcal{F} \left(\beta _b\right)\right)_{ij}=e^{\beta E_i}\delta_{ij}$, and $c(\beta_b)$ a constant that depends only on the bath and choice of normalization for the eigenvector $|v_2\rangle$.
Explicitly using our result for $|v_2\rangle$, we have: 
\begin{eqnarray}
a_2(\beta )=c\left(\beta
   _b\right)\left(1-\frac{e^{\beta _bE_1}
   Z_1\left(\beta _b\right)}{e^{\beta 
   E_1}Z_1(\beta )}\right).
   \end{eqnarray}
   Finally we note that if we assume the spectrum is not completely degenerate, then:
   \begin{eqnarray}
\partial _{\beta }\left(e^{\beta  E_1}Z_1(\beta
   )\right)=-\Sigma
   _{l=2}^n\left(E_l-E_1\right)e^{-\beta 
   \left(E_l-E_1\right)}<0,
   \end{eqnarray}
   since $E_l-E_1>0$ for at least one $l\neq 1$ by the assumption that the spectrum is not completely degenerate. This implies that $a_2(\beta)$ can only vanish at a single point, that is at  $a_2(\beta_b)=0$. In fact, we see that $a_2(\beta)$ is a monotonic function, ruling out both strong and weak Mpemba effects.
   \begin{figure}[htbp]
\begin{center}
\includegraphics[scale=0.5]{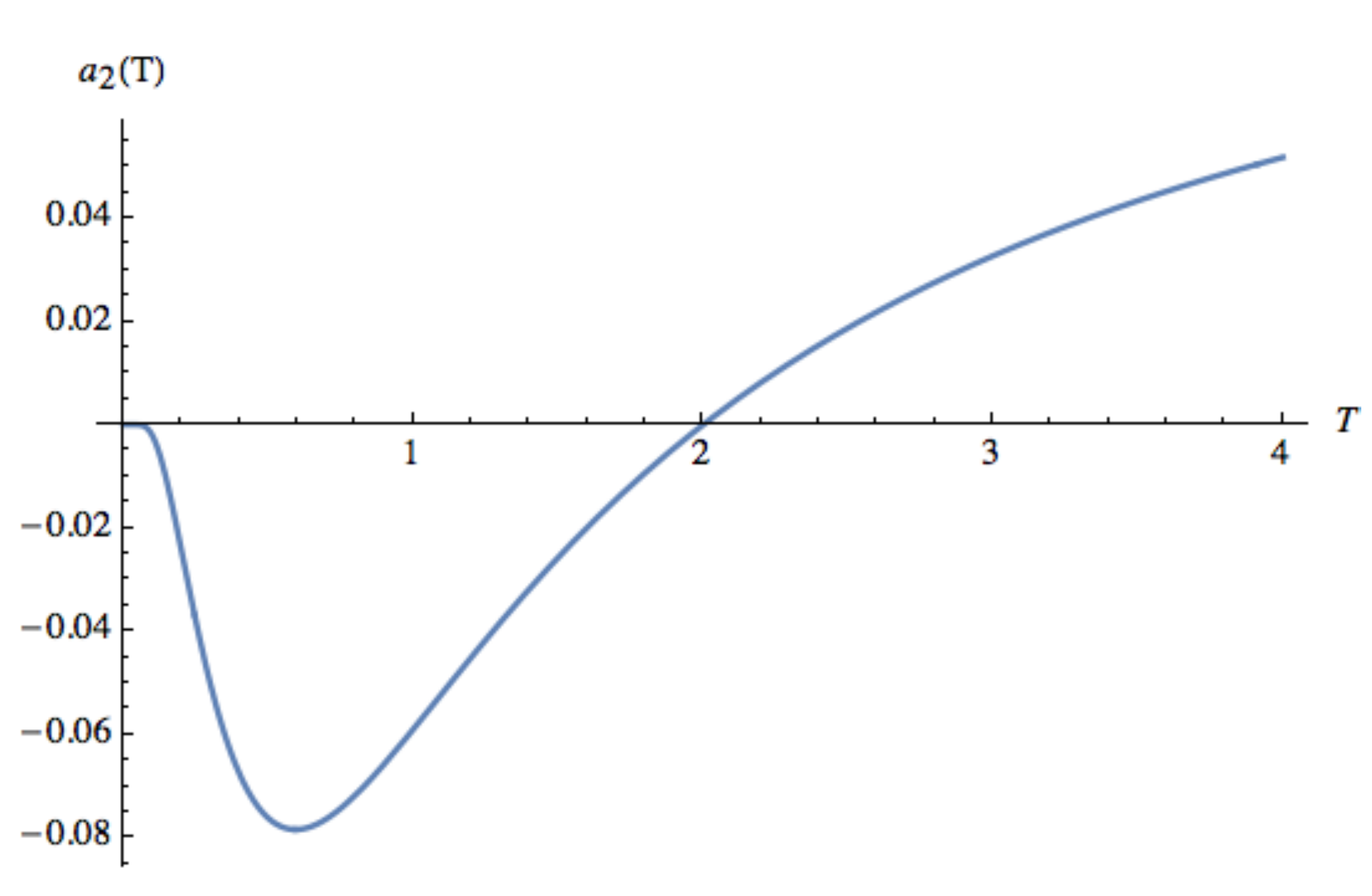}
\caption{ Example of an (indirect) Mpemba effect in a 3 level system with  Metropolis  dynamics, with a single transition blocked. The energies used are $E_1=0,E_2=0.5,E_3=2$ and the bath temperature is set to $T_b=\beta_b^{-1}=2$.}
\end{center}\label{fig:3example}
\end{figure}

\subsection{Mpemba effect with  Metropolis on a graph with a single edge removed}

We now show in a simple example that an Mpemba effect can happen when some of the $R_{ij}$ are set to zero. Let us consider the following $3\times 3$ transition matrix:
\begin{eqnarray}
\left(
\begin{array}{ccc}
-e^{-({E_2}-{E_1})\beta_b}-e^{-({E_3}-{E_1})\beta_b} & 1 & 1
   \\
 e^{-(E_2-E_1)\beta_b} & -1 & 0 \\
 e^{-(E_3-E_1)\beta_b} & 0 & -1 \\
\end{array}
\right)
\end{eqnarray}
This matrix blocks direct $2\leftrightarrow 3$ transitions. The matrix has eigenvalues $0,-1,-1-e^{({E_1}-{E_2})\beta_b}-e^{({E_1}-{E_3})\beta_b}$, with the second eigenvector given simply by:
\begin{eqnarray}
|v_2\rangle =(0,-1,1)
\end{eqnarray}
Computing $a_2$ using \eqref{a2_formula} we have:
\begin{eqnarray}
a_2=\frac{e^{-\left(\beta -\beta _b\right)E_3}-e^{-\left(\beta -\beta _b\right)E_2}}{Z}
\end{eqnarray}
Notice that $a_2(\beta_b)=0$, as should be the case, however, in addition, we see that $\lim_{\beta\rightarrow \infty}a_2(\beta)=0$. Thus between $\beta_b$ and $\beta\rightarrow 0$ (the limit of low temperatures), $a_2$ cannot be monotonic, implying an indirect Mpemba effect, as can be seen in the figure 1.

\section{Discussion}
In this paper, we have shown how the eigenvectors and eigenvalues of the Metropolis dynamics on a complete graph are analytically given when no transitions are blocked. The simplicity of the result is somewhat surprising since the matrix formally depends non-analytically on the energies at energy crossings. As an application we have shown that the Markovian Mpemba effect is not present in this system. The result can be anticipated, as it means that the flow to lower temperatures cannot get "stuck" at an intermediate temperature. On sparser graphs though, the Mpemba effect can arise. Indeed, we give an example of the effect on a simple 3 level example, where one direct transition in the Metropolis dynamics has been set to zero. For larger systems, one can look at the Metropolis simulations of spin-glass in ~\cite{2018arXiv180407569J}, which show the Mpemba effect. 

It is important to note, though, that in most practical applications, i.e., as a numerical tool,  the Metropolis is mostly used on systems with sparse adjacency graphs. On a general graph, an exact solution is not likely. However, the present approach allows for a perturbative treatment when edges of the adjacency graph are removed. 

{\bf Acknowledgement:} The authors thanks Oren Raz for insightful remarks. The work of IK was supported by the NSF grant DMR-1508245. This research was supported in part by the National Science Foundation under Grant No. NSF PHY-1748958.  

\bibliography{Mpemba_MCMC_bib.bib}

\end{document}